\documentclass[final,3p,times,twocolumn]{elsarticle}
\usepackage{graphicx}
\usepackage{amssymb} 
\usepackage{amsmath} 
\usepackage{tabularx}
\usepackage{comment}
\usepackage{multirow}

\usepackage{lineno}

\usepackage{float} 
\usepackage{subfig} 

\usepackage{ifpdf}
\ifpdf
\usepackage[pdftex]{hyperref}
\else
\usepackage[hypertex]{hyperref}
\fi

\hypersetup{
  pdftitle={Physics Opportunities of a Fixed-Target Experiment using the LHC Beams},%
  pdfauthor={S.J. Brodsky, F. Fleuret, C. Hadjidakis, J.P. Lansberg},%
  pdfsubject={},%
  pdfkeywords={},%
  pdfstartview={},%
  bookmarksopen=true, breaklinks=true, debug=true, %
  colorlinks=true, linkcolor=red, citecolor=blue, urlcolor=blue
}

\journal{ArXiv; SLAC-PUB-14878}

\newcommand{\beq}{\begin{eqnarray}}
\newcommand{\eeq}{\end{eqnarray}}
\newcommand{\be}{\begin{eqnarray*}}
\newcommand{\ee}{\end{eqnarray*}}

\newcommand{\cc}{{c\bar{c}}}

\newcommand{\vect}[1]{\vec{#1}}

\newcommand{\eqs}[1]{\begin{equation} \begin{split} #1\end{split} \end{equation} }

\newcommand{\ie}{{\it i.e.}}
\newcommand{\eg}{{\it e.g.}}
\newcommand{\etal}{{\it et al.}}

\renewcommand{\L}{\mathcal L}

\newcommand{\xB}{x_{Bj}}

\newcommand{\cf}[1]{{Fig.~\ref{#1}}}
\newcommand{\ct}[1]{{Table~\ref{#1}}}

\def\lsim{\raise0.3ex\hbox{$<$\kern-0.75em\raise-1.1ex\hbox{$\sim$}}}
\def\gsim{\raise0.3ex\hbox{$>$\kern-0.75em\raise-1.1ex\hbox{$\sim$}}}

\def\jpsi    {\mbox{$J/\psi$}}

\def\beq     {\begin{equation}}
\def\eeq     {\end{equation}}
\newcommand{\sqrtS}[1]{\mbox{$\sqrt{s_{#1}}$}}

\newcommand{\chic}{\mbox{$\chi_c$}}
\newcommand{\chib}{\mbox{$\chi_b$}}

\long\def\symbolfootnote[#1]#2{\begingroup%
  \def\thefootnote{\fnsymbol{footnote}}\footnote[#1]{#2}\endgroup}

\begin{document}


\begin{frontmatter}

\title{Physics Opportunities of a Fixed-Target Experiment using the LHC Beams}

\author{S.J. Brodsky$^1$, F. Fleuret$^2$, C. Hadjidakis$^3$, J.P. Lansberg$^3$}

\address{$^1$SLAC National Accelerator Laboratory, Theoretical Physics, Stanford University, Menlo Park, California 94025, USA \\
$^2$Laboratoire Leprince Ringuet, Ecole polytechnique, CNRS/IN2P3, 91128 Palaiseau, France\\
$^3$IPNO, Universit\'e Paris-Sud, CNRS/IN2P3, 91406 Orsay, France
}

\begin{abstract}
\small
We outline the many physics opportunities offered by a multi-purpose fixed-target experiment
using the proton and lead-ion beams of the LHC extracted by a bent crystal. In a proton run with 
the LHC 7-TeV beam, one 
can analyze $pp$, $pd$ and $pA$ collisions  at  center-of-mass energy $\sqrt{s_{NN}} \simeq 115$ GeV and even 
higher using the Fermi motion of the nucleons in a nuclear target. In a lead run with a $2.76$ 
TeV-per-nucleon beam, $\sqrt{s_{NN}}$ is as high as $72$ GeV. Bent crystals can be used to 
extract about $5\times 10^8$ protons/sec; the integrated luminosity over a year reaches 
0.5 fb$^{-1}$ on a typical 1 cm-long target without nuclear species limitation. 
We emphasize that such an extraction mode
does not alter the performance of the collider experiments at the LHC.
By instrumenting the target-rapidity region, gluon and heavy-quark distributions of the proton and the neutron
can be accessed at large $x$ and even at $x$ larger than unity in the nuclear case.
Single diffractive physics and, for the first time, the large negative-$x_F$ domain can be accessed.
The nuclear target-species versatility provides a unique opportunity 
to study nuclear matter versus the features of
the hot and dense matter formed in heavy-ion collisions, including the formation of the quark-gluon plasma,
which can be studied in $PbA$ collisions over the full range of target-rapidity domain with a large variety of nuclei.
The polarization of hydrogen and nuclear targets allows an ambitious spin program, 
including measurements of the QCD lensing effects which underlie the Sivers single-spin 
asymmetry, the study of transversity distributions and possibly of polarized parton distributions.  
We also emphasize the potential offered by $pA$ ultra-peripheral collisions 
where the nucleus target $A$ is used as a coherent photon source, mimicking photoproduction 
processes in $ep$ collisions. Finally, we note that $W$ and $Z$ bosons  can
be produced and detected in a fixed-target experiment and in their threshold domain for the first time, 
providing new ways to probe the partonic content of the proton and the nucleus.

\end{abstract}


\end{frontmatter}

\tableofcontents


\section{Introduction}
\label{sec:intro}

Fixed-target experiments have played an essential role in hadron and nuclear physics, 
especially in accessing the domain of high Feynman $x_F$ and having the versatility 
of polarized and unpolarized proton and nuclear targets.  Fixed-target experiments  have 
led to the discovery of the $\Omega^-(sss)$~\cite{Barnes:1964pd}, the $J/\psi$~\cite{Aubert:1974js},
the $\Upsilon$~\cite{Herb:1977ek} and the atomic anti-hydrogen~\cite{Baur:1995ck} as well as
evidence for the novel dynamics of quarks and gluons~\cite{Abreu:2000ni} in PbPb collisions. 
Fixed-target experiments have led to the observation of unexpected  QCD phenomena such as the 
breakdown~\cite{Guanziroli:1987rp} of the perturbative QCD Lam-Tung relation~\cite{Lam:1980uc} 
in lepton pair production, novel dynamical effects such as color transparency in diffractive 
dijet production~\cite{Aitala:2000hc}, higher-twist effects in Drell-Yan reactions at high 
$x_F$~\cite{Berger:1979du}, 
anomalously large single~\cite{Adams:1991cs} and double-spin~\cite{Crosbie:1980tb} correlations, 
and the strong non-factorizing nuclear suppression of $J/\psi$ hadroproduction at high 
$x_F$~\cite{Hoyer:1990us}.

The density and length of the target allows fixed-target set-ups  to reach extremely high luminosities.
Thus LHC beams of 7 TeV protons and 2.76 TeV-per-nucleon lead ions interacting on a 
fixed-target\footnote{In the following, we will refer to such a project as "AFTER", standing 
for A Fixed Target ExperiRement \symbol{64} LHC.} 
would provide the opportunity to carry out a large range of precision measurements at unprecedented laboratory energies as well 
as allow the production of a complete range of heavy hadrons such as the $\Omega^-(bbb)$ and exotic states
with a unique access to the large negative-$x_F$ domain.

The collisions of the 7 TeV proton beam on fixed targets correspond to a center-of-mass 
energy close to 115 GeV, half way between those of SPS and RHIC. 
With a nine month per year proton program, one would be able to study the production of quarkonia, open heavy flavor hadrons
and prompt photons in  $pA$ collisions with a statistical accuracy never reached before, especially in 
the target-fragmentation region $x_F \to -1$.  The 
Fermi motion in the nucleus target induces an approximate 10 \% spread of $\sqrt{s}$~\cite{Fredriksson:1975tp}; 
one has indeed the possibility to study the region $x>1$ in detail.
High precision QCD measurements can also obviously be carried out in $pp$ and $pd$ collisions with hydrogen and deuterium
as well as nuclear targets. 

The scheduled Pb-one-month program  at the LHC offers the opportunity to study $\sqrtS{NN}=72$ GeV Pb$A$ collisions
where the Quark-Gluon Plasma (QGP) should be created.
Looking at the QGP in the target rest frame offers the advantage of studying the remnants of the nucleus 
in its rest frame after the formation of the QGP.  The full coverage of the backward region would also allow the
study of long-range near-side correlation in Pb$A$ collisions. 
In addition, thanks to the use of the recent ultra-granular calorimetry technology,
studies of direct photon, \chic\ and even \chib\ production  in heavy-ion collisions 
--two measurements not available in any other experimental configuration-- can be envisioned.
The analysis of Pb$p$ collisions at $x_F \to -1$ using a hydrogen target gives access to small 
$x$ in the Pb nucleus and high $x$ in the proton.

Overall, such a fixed-target facility would  provide a novel testing 
ground for QCD at unprecedented  laboratory energies and momentum transfers as well 
as complete coverage in the rapidity region.   Intrinsic heavy quark distributions at 
large $x_F$ associated with the higher  Fock states of the proton become accessible, 
providing new mechanisms for the production of hadrons with multiple heavy quarks such 
as  baryons with two or three bottom quarks.   A  polarized target would add the possibility 
of studying spin correlations such as the 
non-factorizing~\cite{Sivers:1989cc,Brodsky:2002rv,D'Alesio:2007jt,Barone:2010zz} 
aspects of the Sivers 
effect which pins down the correlation between the parton $k_T$ and the nucleon spin.
A nuclear target could also allow the study of the diffractive dissociation 
of the proton into three jets~\cite{Frankfurt:2002jq}, new tests of color transparency~\cite{Brodsky:1988xz},  
as well as shadowing~\cite{Glauber:1955qq,Gribov:1968jf}
and non-universal antishadowing~\cite{Kovarik:2010uv,Brodsky:2004qa}.  

In addition, the high energies of the LHC beams would render possible the production (and the detection)
of vector bosons such as the $W^+$ for the first time in their threshold region
--thanks to the high luminosity of the fixed-target mode-- and possibly also the $Z^0$. 
Studies in the threshold region could be very important for the search for
heavy partners of gauge bosons, predicted in many extensions to the standard model 
(see \eg~\cite{Hewett:1988xc,Leike:1998wr,Hill:2002ap}) and for which the threshold
region would be reached at the LHC. 

This Letter is organized as follows: In the next section we give the key kinematics
and features of a fixed-target facility of the LHC beams, including luminosities. 
We start with studies related to the proton --and neutron-- partonic structure, 
especially the gluon and the heavy-quark distributions at large $x$. Second,
we detail the opportunity for an ambitious spin 
program allowed by the polarization of the target and measurements of a final-state polarization.
Third, we describe measurements linked to nuclear effects, which are of great interest by themselves, 
but also for QGP studies.
Fourth, we expose the possible analyses of  deconfinement matter in 
Pb$A$ collisions with modern detector technologies and high statistics.
Fifth, we discuss the possibilities offered by the first measurement of $W$ and $Z$ production in
a fixed-target mode.    Finally, we discuss opportunities 
offered by semi-exclusive processes and ultra-peripheral collisions,  which allows a hadron 
collider to be used as a photon-proton/ion collider.
We also briefly enumerate further features offered by a slow extraction of the LHC beam, 
such as high-energy tertiary beams of electrons and secondary beams of hadrons.

\section{Key numbers and features}
\label{sec:key_numbers}

The slow extraction of part of the 7 TeV proton beam from LHC has been investigated 
with the use of a bent crystal~\cite{Uggerhoj:2005ms,Uggerhoj:2005xz} and will be 
experimentally tested within the next two years\footnote{In its september 2011 minutes, 
the LHC Committee recommends that the LUA9 Collaboration~\cite{LUA9}
carry out beam bending experiments using crystals at the LHC.}. Such a device would 
offer the possibility to deflect part of the beam halo at a rate of the order of 
$5\times 10^8$ $p^+$/s without any performance decrease for the LHC collider 
experiments\footnote{The nominal number of protons stored in the LHC ring is of 
the order of $3 \times 10^{14}$ protons ; extracting $5 \times 10^8$ protons/sec for a typical 
10 hours run would reduce the number of protons in the ring by $6\%$.}.

With a similar device, the extraction of Pb ions has been successfully tested at 
SPS~\cite{Scandale:2011zz} and should also be possible at LHC taking advantage, for instance, 
of new techniques to bend diamond crystals~\cite{Balling:2009zz}. We expect that one can achieve
a rate of $2\times 10^5$ Pb/s.\footnote{The nominal number of Pb ions stored in the LHC ring 
is of the order $4.1\times 10^{10}$ ions. Such a rate corresponds to an extraction of about $15\%$ 
of the lead beam over a fill of 10 hours.} 

Tables \ref{tab:lumi-pA} and \ref{tab:lumi-PbA} give the reachable instantaneous luminosities 
obtained with a proton and a Pb beam respectively for various 1cm thick targets as well as the 
integrated luminosities over one year (taken as 10$^7$~s for the proton beam; 10$^6$~s for the Pb beam).
Depending on the target density, the integrated luminosity for the proton beam stands within 0.1 and 0.6 fb$^{-1}$.

\begin{table}[!hbt]\small
\centering\setlength{\arrayrulewidth}{.8pt} \renewcommand{\arraystretch}{0.9}
\begin{tabular}{cccccc}
\hline\hline 
Target       & $\rho$        &$A$ & $\L$                     & $\int\L$\\
(1 cm thick) & (g cm$^{-3}$) &    & ($\mu$b$^{-1}$ s$^{-1}$) & (pb$^{-1}$ yr$^{-1}$)\\
\hline
solid H  & 0.088 & 1   & 26 & 260 \\
liquid H & 0.068 & 1   & 20 & 200 \\
liquid D & 0.16  & 2   & 24 & 240 \\
Be       & 1.85  & 9   & 62 & 620 \\
Cu       & 8.96  & 64  & 42 & 420 \\
W        & 19.1  & 185 & 31 & 310 \\
Pb       & 11.35 & 207 & 16 & 160 \\
\hline\hline 
\end{tabular}
\caption{Instantaneous and yearly luminosities obtained with an extracted beam of 
$5 \times 10^8$ p$^+$/s with a momentum of 7 TeV for various 1cm thick targets}
\label{tab:lumi-pA}
\end{table}
\begin{table}[!hbt]\small
\centering\setlength{\arrayrulewidth}{.8pt}
\renewcommand{\arraystretch}{0.9}
\begin{tabular}{ccccc}
\hline\hline
Target&$\rho$&$A$&$\L$ &$\int\L$ \\
(1 cm thick) &(g cm$^{-3}$)&&(mb$^{-1}$ s$^{-1}$)&(nb$^{-1}$ yr$^{-1}$)\\
\hline 
solid H  & 0.088 & 1   & 11 & 11 \\
liquid H & 0.068 & 1   & 8  & 8  \\
liquid D & 0.16  & 2   & 10 & 10 \\
Be       & 1.85  & 9   & 25 & 25 \\
Cu       & 8.96  & 64  & 17 & 17 \\
W        & 19.1  & 185 & 13 & 13 \\
Pb       & 11.35 & 207 & 7  & 7  \\
\hline\hline 
\end{tabular}
\caption{Instantaneous and yearly luminosities obtained with an extracted beam of 
$2 \times 10^5$ Pb/s with a momentum per nucleon of 2.76 TeV for various 1cm thick targets }
\label{tab:lumi-PbA}
\end{table}

At 7 TeV, a proton beam on a nucleon in a fixed target leads to a center-of-mass energy close
to 115 GeV and a center-of-mass rapidity boosted to 4.8. Translated into the laboratory frame,
the center-of-mass central-rapidity region, $y_{cms}\simeq 0$, is at an angle of 0.9 degrees with respect
to the beam axis. Whereas the backward region ($y_{cms}<0$) can easily be
 accessed with standard experimental techniques, the access to the forward region is 
limited by the distance to the beam axis and would require the use of highly segmented 
detectors to deal with the large particle density. Thus we expect  
most of the measurements to be carried out in the region $-4.8\leq y_{cms}\leq 1$.

Such a rapidity coverage would allow one to detect the bulk of the
particle yields as well as a thorough studies of phenomena in the whole backward hemisphere. 


\section{Nucleon partonic structure}
\label{sec:PDF}

\subsection{Drell-Yan}

The Drell-Yan (DY) process provides a way to access the antiquark content 
of the proton. Indeed, the DY process necessarily involves the antiquark distributions 
of either the beam or target hadron depending on the dilepton
rapidity. By measuring DY pair production with the 7 TeV proton beam of the 
LHC on both hydrogen and deuterium targets in the backward region, we expect 
to access the antiquark distributions, $\bar u (x)$ and $\bar d(x)$, in the nucleons 
at rather low $x$, complementing the forthcoming studies by E906~\cite{Reimer:2011zza}.
In addition, one can test novel QCD effects such as the breakdown of factorization of  the Sivers effect,  the lensing of initial-state interactions~\cite{Boer:2002ju} which via the double-Boer-Mulders  effects produces a large $\cos 2 \phi$ coplanar correlation, and higher twist effects which modify the standard $1 + \cos^2 \theta_{CM}$ 
distribution  at large $|x_F|$  .

From what is known for DY at $\sqrt{s}=200$ GeV (see \cite{Liu-talk}), one expects the DY signal 
to be rather clean of charm and beauty decay background for invariant mass above the 
charmonium family. Studies at lower invariant mass, between 1.5 and 3 GeV where charm 
decay dominates over DY, would require isolating the leptons and maybe also to remove displaced-vertex leptons. 

It will be possible to detect dilepton pairs with invariant masses above 2 GeV  
up to the very backward region, allowing for large momentum fraction in the target 
(large $x_2$). Taking $x_2 \simeq 0.5$, one would reach $x_1 \simeq 4 \times  10^{-3}$ 
for $M_{\ell\ell}=5$ GeV, and value as low as $x_1 \simeq 6 \times 10^{-4}$ for 
$M_{\ell\ell}=2$ GeV. Studies of DY pairs with invariant masses above the bottomonium family
would allow to study larger $x$ parton in the target than 0.5 while keeping the momentum 
fraction in the projectile in a range where the PDFs are very well known~\cite{Holt:2010vj}.

\subsection{Gluons in the proton at large $x$}

Although momentum sum rules tell us that gluons carry about $40\%$ of the proton momentum at $Q^2\simeq 10$ GeV$^2$, 
it is very difficult to probe them directly. Deep-Inelastic Scattering (DIS) experiment
can only directly probe the target quark content. Indirect information on the gluon
content can be extracted from the $Q^2$ dependence of the quark distribution --the scaling violation. 
In DIS and DY at large $x$, the extraction of PDFs is not easy due to the presence of higher-twist
corrections, such as mass effects~\cite{Schienbein:2007gr} and direct processes~\cite{Berger:1979du}. 
Besides, sum rules are of no practical use in this region because of the strong suppression of PDF 
for $x$ approaching 1. As a consequence, the gluon distribution is very badly known for 
$\xB>0.2$ at any scale, as illustrated on \cf{fig:CTEQ10g}.

Obviously, at very large $x$, non-factorizable
contributions could become significant, preventing the gluon PDF extraction. It is 
therefore important to make sure that one is in a region where factorization is 
still tractable. Some non-factorizable effects may occur for all the observables 
such as Sudakov effects. Other, such as coalescence of Intrinsic Charm (IC), 
would matter for quarkonium production at large $|x_F|$. They could be 
studied with a combined analysis of large-$|x_F|$ charm and $J/\psi+D$ production. 
Evidently, close to the edge of the phase space, diffractive processes would start to 
dominate, they are discussed in section \ref{sec:diffraction}. A better understanding
of the interplay between inclusive and exclusive processes is clearly crucial here and calls
 for the possible analyses at a fixed-target experiment on the LHC $p$ beam presented below.

\begin{figure}[thb!]
\centering
\includegraphics[width=0.48\columnwidth]{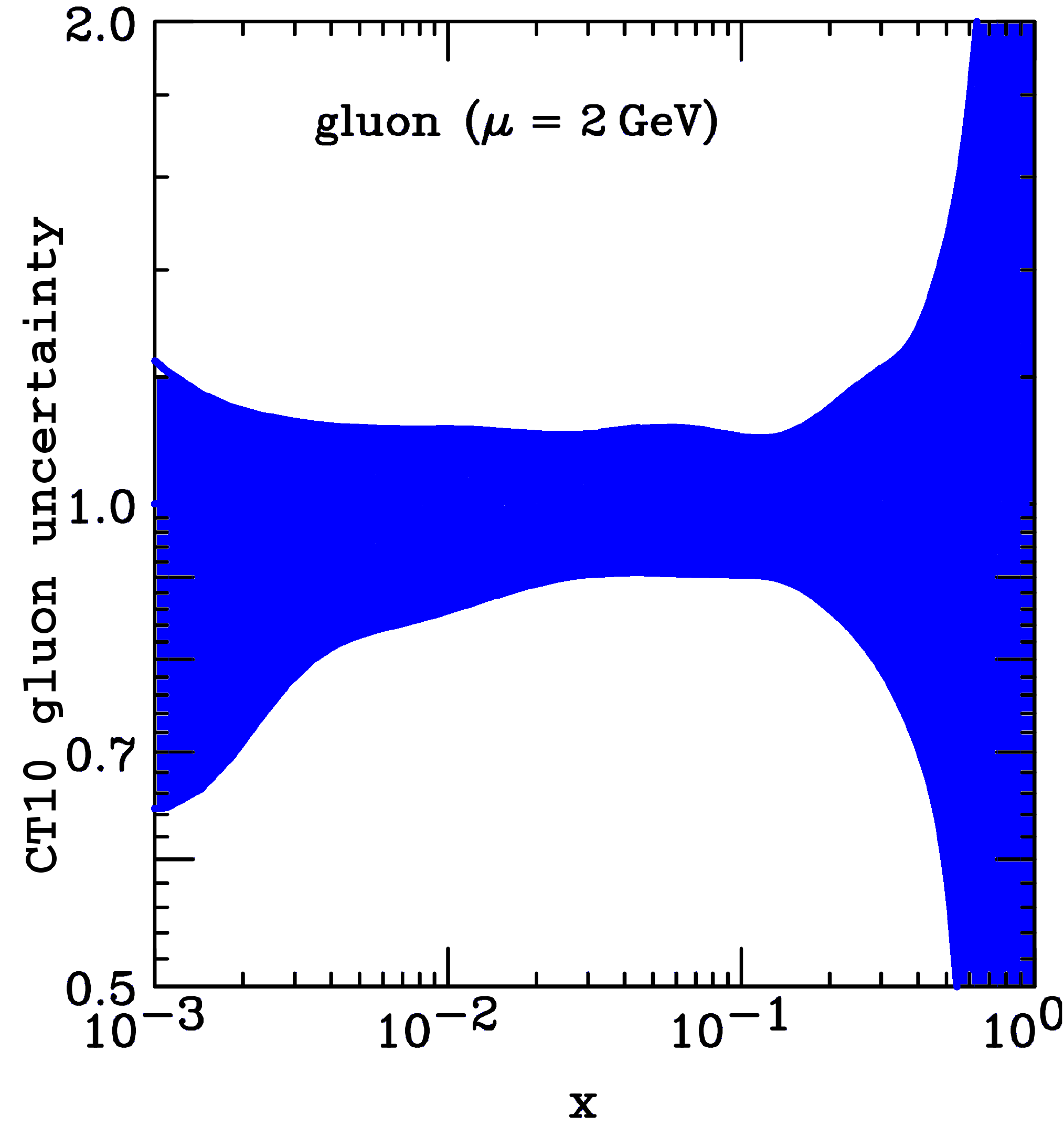}
\includegraphics[width=0.48\columnwidth]{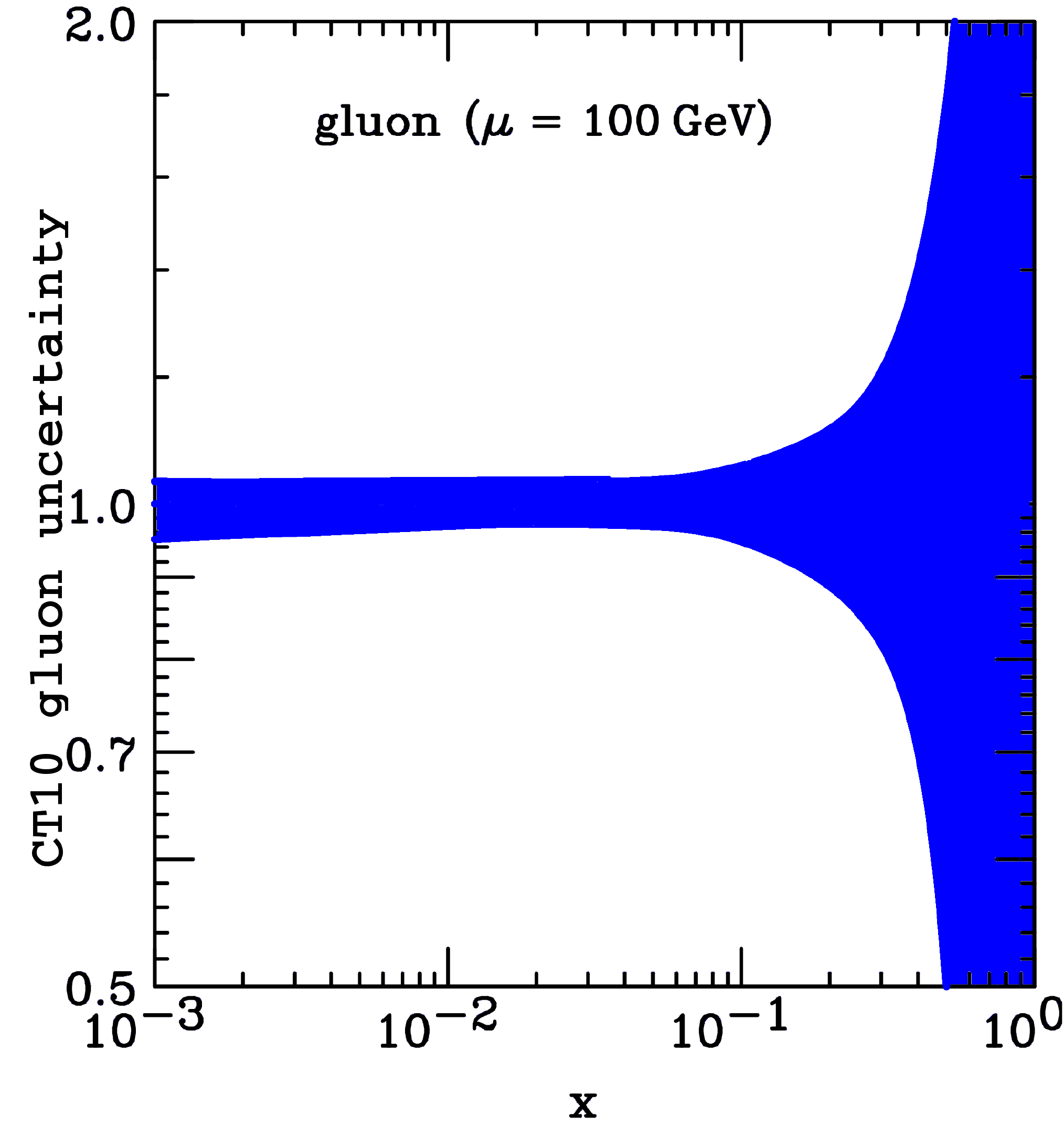}
\caption{Illustration of the relative uncertainties on the gluon distribution at large 
$x$ from the CT10 global analysis~\cite{Lai:2010vv}.}
\label{fig:CTEQ10g}
\end{figure}

\subsubsection{Quarkonia}
\label{subsubsec:onia_pp}

The information which could be obtained from quarkonium production 
in the forward and backward region in LHC laboratory frame experiments should be very valuable. 
Indeed, they are produced by fusion of two gluons at scales commensurate with their mass, thus
large enough to use perturbative QCD (pQCD). Unfortunately, the numerous puzzles --along with the 
large theoretical uncertainties-- in the predictions of $J/\psi$ and $\Upsilon$ production 
rates at hadron colliders\footnote{Let us nevertheless emphasize that the LO prediction for 
the $P_T$-integrated yield of $J/\psi$ from the pQCD-based CSM is in good agreement with the 
RHIC, Tevatron and LHC data without tuning any parameter values~\cite{Lansberg:2010cn}.}  
do not give --at first sight-- a strong incentive to follow up pioneering analyses of gluon 
PDF extraction with quarkonium data~\cite{Martin:1987vw,Glover:1987az}.

On the contrary, the use of $C=+1$ quarkonia\footnote{More specifically $\eta_{c,b}$ 
and $\chi_{c,b}(^3P_2)$.}  should be much more reliable. They are produced at LO without 
a recoiling gluon~\cite{Lansberg:2006dh,Brambilla:2010cs,Maltoni:2004hv}, hence with competitive rates and 
via a Drell-Yan like kinematics where the gluon momentum fractions are directly related to the rapidity
of the quarkonium. Since the leading-$P_T$ scaling is already reached at NLO, large QCD corrections 
as seen for $\psi$ and $\Upsilon$~\cite{Campbell:2007ws,Artoisenet:2007xi,Gong:2008sn,Artoisenet:2008fc} are not expected. 
Finally, it is also worth noting that $\eta_c$ has never been 
studied in inclusive hadroproduction, although a couple of decay channels
are perfectly workable, and its production rate is expected to be as large, if not larger, than that of 
$J/\psi$.

Based on the very large yield --surpassing that of RHIC by 2 orders of magnitude (see \ct{tab:yields})--, 
we can identify two ways to prepare the ground for gluon PDF extraction from quarkonium studies. 
First, one could follow up on the studies which are already performed now at hadron colliders 
(for reviews see \cite{Lansberg:2006dh,Brambilla:2010cs}) with the asset of a very high statistics even 
for the radially excited states ($\psi(2S)$, $\Upsilon(2S)$ and $\Upsilon(3S)$), with very fine rapidity 
and $P_T$ binning, with very accurate feed-down extraction where needed, and finally with a thorough 
analysis of polarization parameters. Whereas this may look as conservative in view of the 
expected improvement from the LHC and RHIC in the next decade, one should emphasize that ATLAS and 
CMS do not have coverage for low $P_T$ $J/\psi$, contrary to ALICE and LHCb. However, ALICE does not 
have for now vertexing capabilities in the forward region and suffers from limited luminosity, hence limited
reach for excited states. On the other hand, LHCb will provide very competitive measurements but only 
in the forward region and not during the heavy-ion runs (see later).

The other and more original way to proceed is to directly study the $C=+1$ quarkonia. With modern, 
ultra-granular electromagnetic calorimeter~\cite{Thomson:2009rp}, one would be able to 
 study both $\chi_{c,b}(^3P_2)$ through $\ell^+ \ell^- \gamma$ decays and $\eta_{c}$ in the $\gamma\gamma$ 
channels. If a specific  effort is brought on Particle IDentification (PID), the study of the $p\bar p$ 
decay channel (see \eg~\cite{Barsuk:2012ic}) is reachable. Doing so, all the hidden charm 
resonance could be studied. Similar studies would also be carried on the bottomonium family.  

Combining these data with forthcoming ones from the LHC at higher energies, the ultimate goal here 
is to put an end to the controversy on quarkonium production. Once this is done, quarkonium 
production in $pp$ collisions could be used as a unique way to extract gluon distribution at small, 
mid and large $x$. 

\begin{table}[!hbt]\small
\centering \setlength{\arrayrulewidth}{.8pt}\renewcommand{\arraystretch}{0.9}
\small
\begin{tabular}{cccc}
\hline\hline
Target       & $\int\!\! dt\L$ & $\left.{\cal B}_{\ell\ell}\frac{dN_{J/\psi}}{dy}\right\vert_{y=0}$&
$\left.{\cal B}_{\ell\ell}\frac{dN_{\Upsilon}}{dy}\right\vert_{y=0}$
\\
\hline
10 cm solid H                & 2.6          & 5.2 10$^7$  & 1.0 10$^5$ \\
10 cm liquid H               & 2            & 4.0 10$^7$  & 8.0 10$^4$ \\
10 cm liquid D               & 2.4          & 9.6 10$^7$  & 1.9 10$^5$ \\
1 cm Be                      & 0.62         & 1.1 10$^8$  & 2.2 10$^5$ \\
1 cm Cu                      & 0.42         & 5.3 10$^8$  & 1.1 10$^6$ \\
1 cm W                       & 0.31         & 1.1 10$^9$  & 2.3 10$^6$ \\
1 cm Pb                      & 0.16         & 6.7 10$^8$  & 1.3 10$^6$ \\
\multirow{2}{*}{$pp$ 
{\scriptsize low $P_T$ LHC (14 TeV)}
\Bigg\{ \!\!\!\!} 
                             & 0.05         & 3.6 10$^7$  & 1.8 10$^5$ \\
                             & 2            & 1.4 10$^9$  & 7.2 10$^6$ \\
$p$Pb {\scriptsize  
LHC (8.8 TeV)}               & 10 $^{-4}$    & 1.0 10$^7$  & 7.5 10$^4$ \\
$pp$ {\scriptsize 
RHIC (200 GeV)}              & 1.2 10$^{-2}$ & 4.8 10$^5$  & 1.2 10$^3$ \\
$d$Au {\scriptsize 
RHIC (200 GeV)}              & 1.5 10$^{-4}$ & 2.4 10$^6$  & 5.9 10$^3$ \\
$d$Au {\scriptsize 
RHIC (62 GeV)}               & 3.8 10$^{-6}$ & 1.2 10$^4$  & 1.8 10$^1$ \\
\hline\hline
\end{tabular}
\caption{ \jpsi\ and $\Upsilon$ inclusive yields per unit of rapidity  expected per 
LHC year with AFTER at mid rapidity with a 7 TeV proton beams on various targets compared to those 
reachable at the LHC in the central region in $pp$
at 14 TeV with the luminosity to be delivered for LHCb and ALICE which have a low $P_T$ \jpsi\
coverage,  in the central region in a typical LHC $p$Pb run at 8.8 TeV and at RHIC\protect\footnotemark~ in $pp$ and
$d$Au collisions at 200 GeV as well as in $d$Au collisions at 62 GeV.
The integrated luminosity is in unit of inverse femtobarn per year, the yields are per LHC/RHIC year.  }\label{tab:yields}
\end{table}

\footnotetext{The luminosity for RHIC are taken from the PHENIX decadal plan~\cite{phenix-decadal}.}

\subsubsection{Jets}
\label{subsubsec:jet}

Measurements of jet production at high transverse momentum are known to provide constraints on
gluon distribution~\cite{Huston:1995tw,Pumplin:2009nk} for $x >0.1$. At the Tevatron,  
$p_T^{jet}$ are of the order of 200-400 GeV thus probing the gluon distribution at very large scale.
Measurements at ISR by UA2~\cite{Alitti:1990aa} probed it at smaller scale, yet above 40 GeV. 
A careful and modern analysis (see \eg~\cite{Salam:2007xv,Cacciari:2008gp}) of jet production at
$\sqrt{s}=115$ GeV for transverse momenta between 20 and 40 GeV would constrain the region $x> 0.1$. 
Jets produced slightly in the backward region would probe larger $x$ for the same $p_T^{jet}$, 
allowing to keep scales as low as 20 GeV. Data at low scales are most important to constrain 
the gluon distribution.

\subsubsection{Direct/isolated photons}
\label{subsubsec:direct_photon}

More than 20 years ago, direct/isolated photons have been recognized~\cite{Aurenche:1988vi} 
as a very promising way to extract gluon PDFs. A global analysis of existing 
world data~\cite{Aurenche:1998gv,Aurenche:2006vj} has however shown that the normalization of most recent fixed-target
data E706~\cite{Apanasevich:1997hm,Apanasevich:2004dr} is visibly higher than that of the other 
fixed-target experiments. The need for low energy data has indeed been recently 
re-emphasized~\cite{d'Enterria:2012yj} in a global analysis incorporating the first LHC data.

Along these lines, a new competitive analysis at low $p_T$ to minimize the scale --preferably using an hydrogen target--
is most welcome as well as studies in the backward region to obtain high-precision data
at large $x_T$. The requirement for a very good detection of $P$-waves quarkonium through the 
channel $\ell^+ \ell^- \gamma$ would directly
promote direct photons measurements as another key option to access gluon distribution in the proton
at a fixed-target experiment on the 7 TeV proton LHC beam.

\subsection{Gluons in the deuteron and in the neutron}

Combined measurements with hydrogen and deuterium target so far provide the most
competitive way to study the partonic structure of the neutron. 
A number of DIS experiments have been carried out with deuterium targets (see~\cite{Airapetian:2011nu} 
for a recent analysis and an exhaustive list of past results). 
As for now, (quark) shadowing in the deuterium  has not been observed.
In 1994, the NA 51 experiment provided~\cite{Baldit:1994jk} 
the first direct evidence from their $pp$ and $pd$ DY data --which are 
directly sensitive to the ratio $\bar u / \bar d$--  for an isospin asymmetry of the quark sea. 

However, studies of the gluon distribution in the neutron, $g_n(x)$, are singularly more complicated.
The NMC collaboration~\cite{Allasia:1990zx} analyzed the production ratio of $J/\psi$ on 
proton and deuterium targets. They found a value compatible with unity, indicative --within their 15\% uncertainty--
that the neutron gluon distribution is similar to that of the proton. More information
were provided by the E866 $\Upsilon$ analysis~\cite{Zhu:2007mja} in $pp$ and $pd$ --with 50 cm-long
$H_2$ and $D_2$ targets-- which confirmed that $g_n(x,Q^2\simeq 100 \hbox{ GeV}^2)\simeq g_p(x,Q^2\simeq 
100 \hbox{ GeV}^2)$ for $0.1\leq x \leq 0.23$.
Such a measurement unfortunately had not been reproduced at Fermilab for $J/\psi$ which would have probed
$g_n(x)$ at lower $Q^2$.

Using a 1m-long  deuterium target, one would obtain (see \ct{tab:yields}) one billion of \jpsi\ and
a million of $\Upsilon$ in one unit of $y$ in $pd$. Such high-precision measurements may allow for the first
measurement of a difference between $g_n(x)$ and $g_p(x)$. In any case, such an analysis
would allow for extraction of $g_n(x)$ in a significantly wider $x$ range and at lower $Q^2$ with $J/\psi$. 
Other gluon-sensitive measurement, such as those discussed in sections \ref{subsubsec:direct_photon} and 
\ref{subsubsec:jet}, could also be carried out.
Finally, we want to stress that we do not anticipate other facilities where $pd$ collisions could be studied 
in the next decade -- at least at high enough luminosities and energies where such measurements could be carried out.

\subsection{Charm and bottom in the proton}
Most parametrizations of the charm and bottom PDFs in the proton 
only have support at low $x$ since it is assumed that they only arise from gluon splitting.  
This assumption may lead to inaccurate predictions.
Recent global QCD analyzes~\cite{Pumplin:2007wg} provide indications that
a non-perturbative intrinsic charm (IC)~\cite{Brodsky:1980pb} can be expected
resulting in charm PDFs larger than conventional fits which
ignore intrinsic heavy-quark fluctuations in the proton.   
Such enhancement agrees with the large-$x$ EMC measurement in $\gamma^\star p \to c X$ ~\cite{Harris:1995jx} 
which has never been repeated.

From the non-Abelian QCD couplings, one expects intrinsic component probability to fall as 
$1/M^2_{Q \bar Q}$~\cite{Brodsky:1984nx,Franz:2000ee}. A similar intrinsic beauty component
is also expected, though smaller.
The heavy-quark pair $Q \bar Q$ in the intrinsic Fock state  is  primarily a color-octet,  
and the ratio of intrinsic charm to intrinsic bottom scales thus scales as $m_c^2/m_b^2 \simeq 1/10$. 
Evidence for the existence of a similar intrinsic light-quark sea in the nucleons in HERMES data~\cite{Airapetian:2008qf} 
has been  lately claimed in~\cite{Chang:2011vx} and it agrees with the $1/M^2_{Q \bar Q}$ scaling.

Careful analyses of the rapidity distribution of open- or hidden- 
charm hadrons in a fixed-target set-up at $\sqrt{s}=115$ GeV are 
be therefore very important especially at backward rapidities to learn more on 
these aspects of QCD.

\subsubsection{Open-charm production}

The measurement of displaced-vertex muon from $D$ decay using modern vertexing technologies 
should provide an effective way 
to extract the rapidity distribution of charm quark and thus 
to discriminate between different models of charm distribution in the proton. Measurements
using the $K+\pi$ decay channel, as done by ALICE~\cite{ALICE:2011aa}, could also be done.

\subsubsection{$J/\psi +D$ meson production}

In \cite{Brodsky:2009cf}, it has been shown that, at $\sqrt{s}=200$ GeV, a significant fraction 
of the $J/\psi$  is expected to be produced in association with a charm quark. It was also emphasized that the 
measurement of the rapidity dependence of such a yield would provide a complementary handle on $c(x)$. 
Such a measurement would be efficiently done by triggering on $J/\psi$ events then by looking for
$D$ as discussed above.

\subsubsection{Heavy-quark plus photon}  

The reaction $\bar p  p \to \gamma c X$  at D$\emptyset$~\cite{Abazov:2009de} at high $P_T$
are sensitive to the charm structure function at light-cone momentum fractions $x_c >0.1$. 
In fact, this measurement is one of the few anomalies in QCD reactions reported at the Tevatron.
It is partly attributable to IC~\cite{Stavreva:2009vi}. At $\sqrt{s}=115$ GeV, the study
of $p  p \to \gamma c X$ with $P_T^{charm}$ above 20 GeV would also be probing the charm 
distribution at $x_c >0.1$. $p  p \to \gamma b X$ could also be studied as a baseline.


\section{Spin physics}
\label{sec:spin}

One of the key assets of a fixed-target experiment is the possibility to polarize the 
target (see \eg~\cite{Crabb:1997cy}) to 
allow for Single Spin Asymmetry (SSA) measurements for various probes\footnote{It has been 
suggested~\cite{Ukhanov:2007zz} that the beam extracted by 
crystal channeling could be polarized. If this can be experimentally verified and shown to 
be sizeable, double-spin-asymmetry measurements should of course be envisioned and well placed
in the physics case of such a fixed-target project.}.
Recently, it has been re-emphasized that a class of parton distribution functions, known as 
“Sivers functions”~\cite{Sivers:1989cc}, may be accessed in SSA for hard-scattering reactions 
involving a transversely polarized proton (see~\cite{D'Alesio:2007jt,Barone:2010zz} for recent reviews). 
These functions express  a correlation between 
the transverse momentum of a parton inside the proton, and the proton-spin vector. As such 
they contain information on orbital motion of partons in the proton. Sivers-type single-spin 
asymmetries have been observed in semi-inclusive DIS (SIDIS) at 
HERMES~\cite{Airapetian:2004tw}  and COMPASS~\cite{Alexakhin:2005iw} as well as
 in single forward $\pi$~\cite{Adams:1991cs,Arsene:2008mi,Abelev:2008qb} and 
$K$ production~\cite{Arsene:2008mi} at Fermilab and Brookhaven.

These SSA are believed to be due to the rescattering of the quarks and gluons in the 
hard-scattering reactions~\cite{Brodsky:2002rv,Collins:2002kn,Brodsky:2002cx}, and in general 
they do not factorize in the standard pattern expected in perturbative QCD. For instance, the 
SSA asymmetries in SIDIS and DY are expected to have opposite signs, though described by the same 
Sivers function. It is therefore very important to measure these SSA for a number of processes. 
Moreover, nearly nothing is known about gluon Sivers functions.

Another mechanism, known as the Collins effect~\cite{Collins:1992kk,Collins:1993kq}, was 
initially expected to be the main source of SSA in single $\pi$ hadroproduction. Recently, 
a careful treatment of the non-collinear partonic interactions showed it to be eventually 
sup\-pressed~\cite{Anselmino:2004ky}. Since it allows the quark transversity, $h_1(x)$, to be probed 
in single polarized collisions, it remains however important to investigate on processes for 
which Collins-type asymmetries may contribute. Both Collins and Sivers asymmetries are believed 
to come from naive T-odd Transverse Momentum Dependent (TMD) effects.

We present below some options where a fixed-target setup at $\sqrt{s}=115$ GeV with high luminosity,
 good coverage in the rapidity region of the transversally polarized-target  (mid and large 
$x_p^\uparrow$), may be extremely competitive and complementary to the other existing high-energy 
particle physics spin projects. 

\subsection{Transverse SSA and DY}

The Drell-Yan process is definitely a key tool to access naive T-odd TMD effects
in the PDF sector and discriminate among them. Starting already from the unpolarized cross
section, we can study $\vect{k}_\perp$ effects in PDFs by studying the $q_T$
spectrum of the final lepton pair~\cite{D'Alesio:2004up}. 

In the corresponding transverse SSA, $pp^\uparrow \to\ell^+\ell^- + X$,
 two TMD mechanisms could play a role: the Sivers effect and the Boer-Mulders effect~\cite{Boer:1997nt}\footnote{This is
the correlation of the quark transverse spin and its transverse momentum, $k_T$, in an unpolarized proton. It explains
the violation of the Lam-Tung relation~\cite{Lam:1980uc} in unpolarized DY reaction.} --also 
involving transversity. As opposed to unpolarized inclusive reactions, such as $pp\to h + X$, the
 dilepton angular distribution analysis allows one to separate out both effects. 
\ct{tab:DY-SSA-projects} shows a comparison of the various luminosities of
various projects where DY SSA could be measured. Clearly, the setup presented
here with a large luminosity and a wide coverage is well placed to measure DY SSA and
the dilepton angular distribution both at low  and large $ x_p^\uparrow$, here $x_2$.

\begin{table}[htb!]\small
\begin{center} \setlength{\arrayrulewidth}{.8pt} \small \renewcommand{\arraystretch}{1}
\begin{tabularx}{\columnwidth}{p{1.5cm}p{0.9cm}p{0.8cm}p{0.5cm}p{1.2cm}p{0.2cm}}
\hline \hline     \small
Experiment  & particles & energy {\small (GeV)} & $\sqrt{s}$ {\small (GeV)} & ~~~$x_p^\uparrow $  & $\L$ {\small(nb$^{-1}$s$^{-1}$)} \\ 
\hline  
AFTER       & $p+p^{\uparrow}$               & 7000     & 115  & $0.01\div 0.9$  &  1    \\             
 \hline  
COMPASS     & $\pi^{\pm}+p^{\uparrow}$        & 160      & 17.4  & $0.2\div 0.3$  &  2    \\              
COMPASS (low mass) & $\pi^{\pm}+p^{\uparrow}$ & 160      & 17.4  & $\sim 0.05$    &  2    \\
RHIC        & $p^{\uparrow}+p$               & collider &  500  & $0.05\div 0.1$ & 0.2   \\ 
J--PARC     &  $p^{\uparrow}+p$              &  50      &   10  & $0.5\div 0.9$  & 1000  \\     
PANDA  
(low mass)  & $\bar{p}+p^{\uparrow}$         &  15      & 5.5   & $0.2\div 0.4$  & 0.2   \\
PAX         & $p^{\uparrow}+\bar{p}$         & collider &  14   &  $0.1\div 0.9$ & 0.002 \\ 
NICA        & $p^{\uparrow}+p$               & collider &  20   &  $0.1\div 0.8$ & 0.001 \\
RHIC Int.Target 1  & $p^{\uparrow}+p$        &   250    &  22   &  $0.2\div 0.5$ & 2     \\
RHIC Int.Target 2  & $p^{\uparrow}+p$        &   250    &  22   &  $0.2\div 0.5$ & 60    \\
\hline      \hline         
\end{tabularx}
\caption{Compilation~\protect\cite{Barone:2010zz,Goto:2010} of the 
relevant\protect\footnotemark ~parameters for the
future planned polarized DY experiments.
 For AFTER, numbers correspond to a 50 cm polarized $H$ target.}
\label{tab:DY-SSA-projects}
\end{center}
\end{table}

Furthermore, it has been emphasized in~\cite{Sissakian:2008th} that studying the unpolarized and 
single-polarized DY processes in the limiting case $x_p \ll  x_p^\uparrow$, one 
can directly extract the ratio of transversity and the first moment of the
Boer-Mulders PDF. This is exactly the easiest limit to look at with a
fixed target on the LHC proton beam, contrary to other fixed-target projects where DY studies 
are planned such as J-PARC and COMPASS where this limit cannot be reach easily.
It is also interesting to check the small size of sea-quark Boer-Mulders function
expected from the  negligible $\cos 2\phi$ dependence observed recently for DY dimuons 
in $pd$ collisions~\cite{Zhu:2006gx}, contrary to $\pi$-induced DY.

Using Sivers TMDs extracted from SIDIS data, DY SSA are predicted to be large at future set-ups where polarized
DY is to be measured~\cite{Anselmino:2009st}. This could allow a clear test of the predicted sign change of the 
Sivers effect in SIDIS and DY. Such fundamental test of our understanding of SSAs, within 
our factorized picture of QCD would definitely be possible with AFTER, especially at 
lower $x$ where SIDIS measurements
exist.

\subsection{Quarkonium and heavy-quark transverse SSA}

Recently, PHENIX has measured~\cite{Adare:2010bd} that the transverse SSA in $p^\uparrow p \to J/\psi X$ deviates
significantly from zero at $x_F\simeq 0.1$ . According the analysis of~\cite{Yuan:2008vn}, this hints at a 
dominance of a color-singlet mechanism at low $P_T$ and at a non-zero gluon Sivers effect. With AFTER, such 
measurements could be extended to larger $x_p^\uparrow$ as well as  to the other charmonium states and to the bottomonium
family. 

Another related SSA is for the open charm, $p^\uparrow p \to D X$, also proposed~\cite{Anselmino:2004nk} as a direct access 
to gluon Sivers effect --provided that IC is not dominant in the studied region. This measurement 
could be carried out using silicon vertex detector in various ways, \eg~tagging 
$\mu$ from $D$ and $B$'s and non-prompt $J/\psi$ from $B$. 
Doing so, one would have a set of observables sensitive to the gluon Sivers effect.

\footnotetext{We nevertheless insist that the integrated luminosities may strongly depend
on the Run duration ($10^7$ s for the LHC). The figure of merit for SSA also strongly depends
on the spin dilution factors.}

\subsection{Transverse SSA and photon}

The study of SSA in $p^\uparrow p \to \gamma X$ was proposed some time ago in~\cite{Qiu:1991wg}
and was shown to be also sensitive to gluon Sivers function~\cite{Schmidt:2005gv} for $\sqrt{s}$ of the
order of 100 GeV. In such a case, the asymmetry may be as large as 10\%~\cite{D'Alesio:2006fp}. Recently, it was proposed
to refine the analysis to look for SSA in photon-jet production~\cite{Bacchetta:2007sz}
with a constraint on the pseudo-rapidities of both
the photon and the jet. For a photon in the polarized-proton rapidity region and for a jet
slightly in the unpolarized proton rapidity region, such SSA is sensitive to the unpolarized
gluon PDFs, $g(x)$, and to quark Sivers effect. In this case, it appears~\cite{Bacchetta:2007sz}
that the generalized parton model (GPM) (see \eg~\cite{D'Alesio:2004up,Anselmino:2004ky}) and the 
color-gauge-invariant QCD formalism~\cite{Qiu:1998ia} predict a different sign for the SSA. 
As such, it is important to measure this asymmetry, which can be done with AFTER 
along with the analysis of the unpolarized cross section discussed in the previous sections.

\subsection{Spin Asymmetries with a final state polarization}
 
Hyperon ($\Lambda$, $\Sigma$, ...) production in single polarized $p^\uparrow p$ collisions  is also known as a promising
way to access transversity~\cite{deFlorian:1998am}. The measurement of the spin-transfer asymmetry, $D_{NN}$, between
the initial polarized proton and the hyperon involves not only the transversity distributions in the proton
but also the corresponding hyperon transversity fragmentation functions which can be measured at
$e^+e^-$ colliders. At $\sqrt{s}=19$ GeV, such asymmetry was found very large for 
$\Lambda$~\cite{Bravar:1997fb}. It is very important to extend such analysis to larger
 energies and to other hyperons. This can be done with
AFTER, even at larger $P_T$ to check that it is a leading twist (transversity) or higher-twist effect.

Moreover, some time ago, the extraction of $\Delta g$ with a sole polarized target 
has been proposed~\cite{Cortes:1988ww} by measuring the helicity of 
the $\chi_{c2}$ via its decay intro $J/\psi+\gamma$. Such a kind of 
measurement can be envisioned with AFTER.


\section{Nuclear matter}
\label{sec:nuclear_matter}

A number of effects characteristic of nuclear matter can be investigated
in precision measurements of hard processes in $pA$ collisions. The first
is the modification of the partonic densities inside bound nucleons.
At very large $x$, Fermi motion is known to modify the PDFs. 
For $0.3 \leq  x \leq 0.7$, a depletion of the PDFs is observed, but
there is no consensus on the physical origin of this  --EMC-- effect.
Antishadowing --an excess of partons compared to free nucleons at mid $x$-- is for instance 
present in electron-nucleus deep inelastic reactions, but appears to 
be absent in the case of Drell-Yan processes in $pA$ and neutrino charge current 
reactions~\cite{Kovarik:2010uv}.  One possibility is that antishadowing is quark or 
antiquark specific because of the flavor dependence of Regge exchange in the diffractive 
physics underlying Glauber scattering~\cite{Brodsky:2004qa,Brodsky:1989qz}. Another 
possibility is it is higher twist. At small $x$, below say 0.05, 
the PDFs of nucleons pertaining to nuclei are depleted compared to free ones. This
is referred to as nuclear shadowing~\cite{Glauber:1955qq,Gribov:1968jf} 
expected from the Lorentz contraction of the nucleus at high-energies and thus from
the overlap of the nucleons. 
Other very interesting QCD effects are also
at play such as energy loss of parton, color-screening of Intrinsic Charm (IC), Sudakov suppression
specific to nuclear reactions, etc.  A high luminosity fixed-target experiment
with versatile target choice is the best set-up to explore this physics. Whereas it is 
of significant importance to interpret the physics of hard scatterings 
in $AA$ collisions discussed in section \ref{sec:deconfinement}, this physics is 
genuinely at the small distance interface between particle 
and nuclear physics.

\subsection{Quark nPDF: Drell-Yan in $pA$ and Pb$p$}

As we explained for $pp$ collisions, the Drell-Yan reaction can be competitively
studied with AFTER. In $pA$ collisions, one can access large-$x$
quark distribution in the nucleus target. These are known to be affected by
anti-shadowing, EMC effect and then Fermi motion. Careful investigations
of the $A$ dependence of these effects could easily be achieved 
by changing the target.

In the more original reverse mode, lead on hydrogen (Pb$p$), one would be able to study
antiquark distribution in the lead projectile from rather low $x$ ( $\simeq 5 \times 10^{-3}$)
to larger ones by scanning the invariant mass of the dilepton $Q^2$ to 
larger values.

\subsection{Gluon nPDF}

Whereas gluon shadowing in the existing nPDF constrained fits, especially 
in EPS 08~\cite{Eskola:2008ca} \& 09~\cite{Eskola:2009uj}, is the subject of intense on-going debates, 
the gluon EMC suppression 
is usually overlooked. Indeed, very little is known about gluons in this region and few data 
constrain their distribution at $x$ larger than 0.3. The amount of the EMC suppression is 
actually pretty much unknown~\cite{Eskola:2009uj}, except for a loose constraint set by 
momentum conservation.  Measurement of low $x$ gluon shadowing is one of the flagships of 
Electron-ion projects (eRHIC, ELIC, LHeC). 

On the other hand, access to larger $x$ gluon nuclear modification, especially at low scales, 
will be difficult. This is where $pA$ measurements, such as the ones detailed below, are competitive.

\subsubsection{Isolated photons and photon-jet correlations}

Isolated photon studies appears to be~\cite{Arleo:2007js} to be a promising channel 
which allows for a reliable extraction of the gluon density, $g^A(x)/g^p(x)$, and the structure 
function, $F^A_2(x)/F^p_2(x)$, in a nucleus over that in a proton.

 Looking for prompt photons in $pA$ in the backward hemisphere is {\it a priori} not the most favorable 
case since we would probe gluons at low $x$ --the realm of EIC projects-- in the proton and 
valence quark at larger $x$ in the nucleus target A. Let us however stress that the requirement for large-$P_T$
photon about $y_{cms}=0$ --as done by the fixed-target experiment E706~\cite{Apanasevich:2005gs}-- would allow one to probe 
large-$x$ gluon in the target, which is more interesting.  
The Pb$p$ collisions in the backward region would probe gluons with smaller $x$ in the Pb
projectile. The most promising analysis would certainly be done by looking at photon-jet correlations
which have the virtue of providing more information on the momentum fractions involved in the reaction. Obviously
these measurements can be complemented by single jet production in $pA$ as proposed for gluon studies in $pp$.

\subsubsection{Precision quarkonium and heavy-flavour studies}

Although RHIC experiments pioneered in extending quarkonium studies in $p(d)$Au collisions above 
the 100 GeV limit and in providing insights that gluons are indeed shadowed at low $x$ in the Au 
nucleus~\cite{Adler:2005ph}, they are limited by luminosity constraints typical of colliders. For instance, the existing 
PHENIX $J/\psi$ data are not precise enough to distinguish between $2\to 1$ and $2\to 2$ 
production mechanism~\cite{Rakotozafindrabe:2010su} and no $\psi'$ yield has been measured
so far in $d$Au collisions. In comparison, fixed-target data in $pA$ at lower energy from E866
were precise enough to uncover a difference of the absorption between $J/\psi$ and $\psi'$ at low $x_F$ where
they are formed before escaping the target nucleus~\cite{Leitch:1999ea}. Unfortunately, such data by E866 does not
exist for the bottomonium family. Some initial hints of a strong gluon EMC effect have been 
found in $\Upsilon$ production at RHIC~\cite{Ferreiro:2011xy}, but more precise data are clearly awaited for.

In this context, the large yield we expect both for charmonium and bottomonium production at 
$\sqrt{s_{NN}}=115$ GeV, $10^9$ $J/\psi$ and $10^6$ $\Upsilon$ per year and per unit of rapidity 
(see \ct{tab:yields}) should allow for a very precise study of their production in $pA$ collisions. 
An important point to keep in mind is the  versatility of the target choice. This is a strong asset
to investigate the dependence on the impact-parameter, $\vect b$, dependence of nuclear-matter effects, 
in particular that of the nPDFs~\cite{Klein:2003dj}. The precision and the interpretation of the RHIC studies 
--using the sole $d$Au system-- 
is indeed limited by the understanding and the measurements of the so-called  centrality classes.

With a good enough resolution one can measure ratios of yields such
as $N_{J/\psi}/N_{\psi'}$. With  vertexing, we would have access to open charm and beauty, allowing the measurements of other
ratios such as  $N_{J/\psi}/N_D$ and $N_{\Upsilon}/N_B$ where the nPDF effect may cancel. With a 
good photon calorimetry, we would be able to carry out a systematic study of $\chi_c$ and $\chi_b$, extending 
those of HERA-B \cite{Abt:2008ed}. Of course, $\eta_c$ would also be studied for the first time in $pA$ collisions. 
A combined analysis of these observables would certainly put stringent constraints on the gluon 
distribution in nuclei at mid and large $x$, given that they would also help understand other effects 
at work as we detail now.

\subsection{Color filtering, energy loss, Sudakov suppression and hadron break-up in the nucleus}

For large negative or positive $x_F$, IC may be the dominant source of charmonium production. However
as discussed above, the IC Fock state has a  dominant color-octet 
structure: $\vert(uud)_{8C} (c \bar c)_{8C}\rangle$. In $pA$ collisions at large positive $x_F$,  the color 
octet $c \bar c$ comes from the proton and converts to a color singlet by gluon exchange 
on the front surface of a nuclear target and then coalesces to a $J/\psi$ which interacts 
weakly through the nuclear volume~\cite{Brodsky:2006wb}. One then expects a  $A^{2/3}$ dependence of the rate
 corresponding to the area of the front surface in addition to the $A^1$ contribution from the 
usual pQCD contribution. This is consistent with charmonium production 
observed by the CERN-NA3 ~\cite{Badier:1981ci} and the Fermilab E866 collaborations~\cite{Leitch:1999ea}.

Because of these two components, the cross section violates perturbative QCD factorization for hard 
inclusive reactions~\cite{Hoyer:1990us}. Other factorization-breaking effects exist such
as Sudakov suppression induced by the reduced phase space for gluon emission at large $x_F$,
fractional energy loss, etc. They all deserve careful analyses.

For negative $x_F$, the IC emerges from the nucleus and is thus potentially subject
to nuclear modifications similar to anti-shadowing, EMC or Fermi motion. One does not
expect color filtering anymore. In this rapidity region, the mesons are also fully formed when escaping the nucleus.
The survival probability to do so, usually parametrized by an effective cross section,
  is minimal and related to their physical size. 
A fine study of yield ratios of different quarkonia at negative $x_F$ would 
be very instructive on this matter (see \eg~\cite{Koudela:2003yd}). This region is also
not expected to be affected by fractional energy loss~\cite{Arleo:2010rb}. In the case
of $\Upsilon$, the EMC effect may happen to be the only visible effect at 
work at negative $x_F$ (see~\cite{Ferreiro:2011xy}).
A scan in $x_F$ would also be helpful as an attempt to study the 
$\sqrt{s_{\psi N}}$ dependence of this effective break-up cross section,
which may be non-trivial due to higher twist effects~\cite{Kopeliovich:2001ee}
 
All of these aspects can be investigated with DY, quarkonium, prompt photon and heavy-flavour measurement
at AFTER. It has to be noted that HERA-B is so far the only experiment which could easily access the
region of negative $x_F$. Its reach was nevertheless bound to -0.3 for $J/\psi$~\cite{Abt:2008ya} for instance. So far, 
no other facility could ever go  below that.


\section{Deconfinement in heavy ion collisions}
\label{sec:deconfinement}

Thanks to its energy in the center-of-mass of the collision of $\sqrtS{NN}=72$ GeV in 
Pb$A$ collisions and of $\sqrtS{NN}=115$ GeV in $pA$ collisions, such a fixed target experiment would be very 
well placed to participate to the study of the quark-gluon plasma formation in heavy ion collisions.
Among the large variety of proposed observables, we note, at first sight, that probes such as 
quarkonium suppression, jet quenching or direct photons could be easily accessed with our foreseen 
apparatus. Nuclear Matter effects studies as needed baseline also enter this scope.
Some keys studies are detailed below.

\subsection{Quarkonium studies}

Since the first prediction of \jpsi\ suppression as a probe of the QGP~\cite{Matsui:1986dk}, important results 
have been obtained in PbPb collisions at \sqrtS{NN}=17 GeV at CERN/SPS~\cite{Abreu:1997jh,Abreu:1999qw,Abreu:2000ni}, 
in AuAu collisions at \sqrtS{NN}=200 
GeV at BNL/RHIC~\cite{Adare:2006ns,Adare:2008sh} and recently in PbPb collisions at \sqrtS{NN}=2.76 TeV at 
CERN/LHC~\cite{Abelev:2012rv,Chatrchyan:2012np,Aad:2010px}. 
These results tend to indicate, as it was predicted, that the \jpsi\ production cross section 
is modified by the hot and dense matter which is
produced, but a definite and precise description of these effects on \jpsi\ production is still not at hand. 

\begin{table}[!hbt]\small
\centering \setlength{\arrayrulewidth}{.8pt}\renewcommand{\arraystretch}{0.9}
\begin{tabular}{cccc}
\hline\hline
Target       & $\int\!\! dt\L$ & $\left.{\cal B}_{\ell\ell}\frac{dN_{J/\psi}}{dy}\right\vert_{y=0}$&
$\left.{\cal B}_{\ell\ell}\frac{dN_{\Upsilon}}{dy}\right\vert_{y=0}$
\\
\hline
10 cm solid H       & 110   & 4.3 10$^5$   & 8.9 10$^2$ \\
10 cm liquid H      & 83   & 3.4 10$^5$   & 6.9 10$^2$ \\
10 cm liquid D      & 100   & 8.0 10$^5$   & 1.6 10$^3$ \\
1 cm Be             & 25    & 9.1 10$^5$   & 1.9 10$^3$ \\
1 cm Cu             & 17    & 4.3 10$^6$   & 0.9 10$^3$ \\
1 cm W              & 13    & 9.7 10$^6$   & 1.9 10$^4$ \\
1 cm Pb             & 7    & 5.7 10$^6$   & 1.1 10$^4$ \\
$d$Au {\scriptsize 
RHIC (200 GeV)}     & 150   & 2.4 10$^6$   & 5.9 10$^3$ \\
$d$Au {\scriptsize 
RHIC (62 GeV)  }    & 3.8   & 1.2 10$^4$   & 1.8 10$^1$ \\
AuAu {\scriptsize 
RHIC (200 GeV)}     & 2.8   & 4.4 10$^6$   & 1.1 10$^4$ \\
AuAu {\scriptsize 
RHIC (62 GeV)}      & 0.13  & 4.0 10$^4$   & 6.1 10$^1$ \\
$p$Pb {\scriptsize 
LHC (8.8 TeV) }     & 100   & 1.0 10$^7$   & 7.5 10$^4$ \\ 
PbPb {\scriptsize 
LHC (5.5 TeV) }     & 0.5   & 7.3 10$^6$   & 3.6 10$^4$  \\
\hline\hline
\end{tabular}
\caption{ \jpsi\ and $\Upsilon$ inclusive yields per unit of rapidity  expected per LHC year with AFTER at mid rapidity 
with a 2.76 TeV lead beam on various targets compared to the projected nominal yield in Pb$p$ and PbPb runs
of the LHC at 8.8 and 5.5 TeV as well as in $d$Au and AuAu collisions at 200 GeV and 62 GeV at RHIC. 
The integrated luminosity is in unit of inverse nanobarn per year, the yields 
are per LHC/RHIC year.  }\label{tab:yieldsPb}
\end{table}

\ct{tab:yieldsPb} displays the expected\footnote{For a fair comparison, all these numbers hold for
a single unit of $y$, with the branching into dileptons, without any reduction due to nuclear 
effects, without taking into account the measurement efficiencies and were obtained from 
extra/interpolated cross sections of inclusive yields.} yields for \jpsi\ and $\Upsilon$ using 
the 2.76 TeV Pb beam on various targets. They are compared to those expected nominally per year at 
RHIC in $d$Au and AuAu (at $\sqrtS{NN}=62$ and 200 GeV), at the LHC in Pb$p$ (at \sqrtS{NN}$=8.8$ TeV) 
and  in PbPb (at $\sqrtS{NN} =5.5$ TeV). 

As regards $AA$ collisions, one sees that the yields in PbPb (at $\sqrtS{NN}=72$ GeV) are about equal to 
(100 times larger than) those expected in a year at RHIC for AuAu at $\sqrtS{NN}=200$ GeV 
(62 GeV) and also similar to that to be obtained 
during one LHC PbPb run, despite the lower cross section at lower energies. The same global picture also applies for 
other quarkonium states --as well as for most of the hard probes for QGP studies.

For $pA$ collisions, the inverse PbH mode (still at $\sqrtS{NN}=72$ GeV) 
with a 10cm thick H target offers yields 450 times lower that in the normal $p$Pb mode 
at $\sqrtS{NN}=115$ GeV (see \ct{tab:yields}). However, these 
are only about 20 times less than of LHC $p$Pb mode at $\sqrtS{NN}=8.8$ GeV  
and one sixth of (35 times) that at RHIC in $d$Au at 200 GeV (62 GeV).
A 100cm $H$ as done by NA 51 would allow for 10 times larger yields.

By using novel ultra-granular calorimetry techniques, one would be able to study other
 charmonium states such as 
\chic\ (which can be studied in its $\jpsi+\gamma$ decay channel), thus giving new constraints towards the 
understanding of quarkonium anomalous suppression. It is interesting to note that, at $\sqrtS{NN}=72$ GeV, 
the number of $\cc$ pairs produced per collisions should small enough so that, for any scenario of 
recombination process~\cite{Andronic:2003zv}, we could neglect its effect.

\subsection{Jet quenching}
The suppressed hadron production at large transverse momentum, the so-called jet quenching, has been
observed for the first time at RHIC in central AuAu collisions~\cite{Adcox:2001jp,Adler:2002tq}. This 
suppression, interpreted as a clear sign of the production of a new state of matter, has also been 
observed at $\sqrtS{NN}=62$ GeV, while no such effect has been seen at $\sqrtS{NN}=22.4$ GeV~\cite{Adare:2008cx}. 
With good momentum and energy resolutions, detailed study of both neutral and charged hadrons suppression can
 be performed. The use of a specific PID detector would in addition give the capability to study the
various charged hadron species which suffer jet quenching.

\subsection{Direct photon}
At RHIC, a direct photon excess has been observed in AuAu collisions at \sqrtS{NN}=200 GeV via 
the measurement of low mass $e^+e^-$ pairs~\cite{Adare:2008fqa}. This study led to a measured temperature $T\simeq220$ MeV; 
this is well beyond the temperature where the phase transition
occurs as predicted by lattice QCD calculations. 
Such a measurement at lower energy could give a very useful information on the evolution of temperature 
as a function of the center of mass energy and is accessible 
with a good tracking resolution and good electromagnetic calorimeter granularity.

\subsection{Deconfinement and the target rest frame} 

The production of the deconfined phase of QCD can be explored in the nuclear target 
rest frame in heavy-ion nuclear target collisions at AFTER.  In particular, one can study the 
remnants of the nucleus in its rest frame after the formation of the QGP.

The property of extended longitudinal scaling observed by Phobos~\cite{Back:2004je},
\ie~the energy independence of the charged particle pseudo-rapidity density and of the elliptic flow
over a broad pseudo-rapidity range, when effectively viewed --by boosting the distribution-- in the rest frame of one of the 
colliding nuclei, could be studied effectively by analyzing particle multiplicities and asymmetries 
in the target-nucleus region. In the laboratory frame, the production process of the soft particles 
should indeed be independent of the energy or rapidity of the other particle.

A large rapidity coverage is also be a key asset for 
a precise study of long-range near-side angular correlation, also known as the ridge, 
in $AA$ collisions as done at RHIC and the LHC (see \eg~\cite{Alver:2009id} and \cite{Chatrchyan:2011eka}).

\subsection{Nuclear-matter baseline}

For a reliable extraction of the effects attributed to deconfinement in heavy ion collisions,
the effects on the used probe due to nuclear matter intrinsic to both ions colliding 
need to be subtracted. These can be measured with $pA$ collisions where the 
nuclear effects on the probe from the nucleus $A$ can be isolated from that of the deconfined matter.

The $pA$ program at $\sqrtS{NN}=115$ GeV we proposed 
in section~\ref{sec:nuclear_matter} is very well placed to complement and greatly extend
the studies already made at RHIC and Fermilab. Because of the very high boost, the full backward
 rapidity region would be conveniently accessible, opening the entire negative 
$x_F$ region to a number of analyses. One thus expects to precisely quantify  any nuclear 
effect occurring in this region. This is supported by the capability to perform 
extensive $A$-dependence studies with very large statistical samples, which 
would provide a unique direct survey of the dependence of these effects at both low (light systems) 
and high nuclear densities (heavy systems). Finally, in the inverse kinematical domain, that is 
with Pb$p$ collisions, at $\sqrtS{NN}=72$ GeV, the entire positive $x_F$ region becomes reachable. 


\section{$W$ and $Z$ boson production in $pp$, $pd$ and $pA$ collisions}
\label{sec:W_Z}

$W$ and $Z$ production can provide definitive probes of quark distributions at large $x$, 
including the EMC and Fermi-motion regions. At $\sqrt{s}=115$ GeV, for a $Z$ boson produced 
with $y_Z=0$, $x_{1,2}= M_Z/\sqrt{s} e^{\pm y_Z} \simeq 0.8$. The counterpart of being so close to 
threshold ($x\to 1$) is the small size of the production cross section and that threshold effects
may be sizeable. At LO, we have
\eqs{
\sigma^{pp\to W^+\to \mu^+ \bar \nu_\mu}(|y^\mu|<1,p_T^\mu \geq 5\hbox{ GeV}) \simeq 80 \hbox{fb} \\
\sigma^{pp\to Z\to \mu^+ \mu^-} (|y^\mu|<1,p_T^\mu \geq 5\hbox{ GeV}) \simeq 4 \hbox{fb}
}
NLO and NNLO corrections to the yield~\cite{Hamberg:1990np} and the rapidity 
distributions~\cite{Anastasiou:2003yy} are expected to be important (maybe as large as five times the 
LO itself~\cite{PC-Vogelsang}) and threshold resummation has to be carried out to obtain a reliable 
evaluation of the yield. In any case, the final result would be strongly dependent on the quark PDFs 
at large $x$. Indeed, $W$ and $Z$ boson production
at $\sqrt{s}=115$ GeV would be a crucial discriminate of PDF sets.

\subsection{First measurements in $pA$}

Despite the small cross-section mentioned above, we can take advantage of a nuclear target 
with large atomic number $A$ to achieve reasonably large counting rates, the cross section 
in $pp$ being multiplied by $A$. With an integrated luminosity of 0.3 fb$^{-1}$ on a 1cm W 
target (A=180), one may expect 4000  $W^+$ events with a detector with an acceptance of 2
units of rapidity about 0. Along the same lines, we would detect  about 200 $Z^0$. The 
natural scale of the process, $Q\simeq M_{W,Z}$, being quite large, no coherent effect 
(\ie~shadowing) is expected. Only Fermi motion in the target will be at work and will 
affect the result but towards higher values.

If the measurements are successful and the yield enables it, studies more backward 
than $y_Z=\ln(M_Z/\sqrt{s})=-0.22$ allow one to probe quark distribution in the target 
for $x$ larger than one. We would then reach a {\it terra incognito} between
particle and nuclear physics, by probing a extremely small size parton, of the order 
of a thousandth of a femtometer, whose dynamics is governed by the interaction between 
the nucleons with the nucleus.

\subsection{$W/Z$ production in $pp$ and $pd$}

According to our rudimentary LO evaluation, one may expect a couple of thousands of events 
in $pp$ collisions per year. If needed, the number can be increased by taking a longer target 
--remember that NA51 used a 1.2m long H and D target~\cite{Baldit:1994jk,Abreu:1998rx}. Using 
hydrogen target has the critical advantage of not being sensitive to Fermi motion and thus 
to provide a unique handle on PDFs at large $x$ and/or threshold effects which may otherwise 
be crucial if heavy partners of gauge bosons were discovered at the LHC.

Obviously the target can be polarized, which would allow one to extract information on $\Delta q$ and  
$\Delta \bar q$~\cite{Nadolsky:2003ga}. We expect such measurements to be absolutely complementary to those (to be) 
obtained at RHIC~\cite{Adare:2010xa,Aggarwal:2010vc}.

Beyond these opportunities to study the nucleon structure, studies of $W$ and $Z$ so close to threshold
offer the possibilities of detailed studies of their decay product in a clean environment. Let us 
mention the decay of $Z$ in two jets carrying
80\% of the center-of-mass energy.

\section{Exclusive, semi-exclusive and backward reactions}
\label{sec:diffraction}

\subsection{Ultra-peripheral collisions}

The study of events where two grazing nuclei (or even nucleons) interact electromagnetically, 
namely Ultra Peripheral Collisions (UPCs), effectively turns heavy-ion colliders into photon colliders (for reviews
see~\cite{Baur:2001jj,Bertulani:2005ru}). The key point here is that the strong electromagnetic fields
of the nuclei ($\propto Z^2$) are highly boosted and allows for hadronic systems to be produced via 2 $\gamma$ exchanges,
via $\gamma$ and pomeron exchanges or via single $\gamma$ exchange with a nucleon dissociation. Experimental 
proofs of principle were provided at RHIC in AuAu collisions by the measurement of 
coherent $\rho^0$ production~\cite{Adler:2002sc} and then of
coherent $J/\psi$ production~\cite{Afanasiev:2009hy}. 

The study of UPC at the LHC in $pp$, $pA$ and $AA$ collisions has attracted a 
lot of interest \cite{Baltz:2007kq} in the recent years. 
For instance, it was shown~\cite{Strikman:2005yv} that UPCs in $pA$ and $AA$ collisions would extend
the coverage of HERA for nuclear and gluon PDFs. Nevertheless, one of the main issue to face 
in $pp$ and $pA$ runs is the important pile-up (see~\eg~\cite{d'Enterria:2009er}). With the slow
extraction from a bent crystal as described in section~\ref{sec:key_numbers}, we have evaluated that pile-up 
is absent for hydrogen targets and about one for a typical 1cm-thick
lead target. Such a number is certainly unproblematic.
At $\sqrt{s_{NN}}=115$ GeV, the kinematical range of $pA$ UPCs is obviously reduced, although the $\gamma p$ 
invariant mass,$W_{\gamma p}$,  can be as large
as 40 GeV for coherent processes. Yet, the absence of pile-up thanks to the slow extraction offers many possibilities. 
Vector-meson elastic and inelastic photoproduction could be studied along the lines of HERA studies, more
particularly $pA \overset{\gamma}{\to} (X)~\psi(2S)+X~(A)$ which could not be carried out by H1 and ZEUS
due to resolution limitation. Extending DY measurement in $pA$, one could
study timelike DVCS, $pA \overset{\gamma}{\to} (p) ~\ell^+\ell^-~ (A)$, aiming at the extraction of GPDs~\cite{Pire:2008ea}.

\subsection{Hard diffractive reactions}  

A substantial fraction of the total $pp$ cross section is due to  
single and double diffractive reactions such as $ p p \to X~(p^\prime) $ and $ p p  \to  (p^\prime)~X~(p^\prime)$,
where $X$ is a produced massive state and $p^\prime$ are either protons or low-mass systems
with energies close that of the colliding particles. In such reactions, the final states
particle belong to ``clusters'' with large rapidity gap between them. They 
are sensitive to the diffractive PDFs (DPDFs)~\cite{Martin:2006td,Aktas:2006hy,Chekanov:2009qja} 
of the proton and can be described by Regge theory~\cite{Collins:1977jy} with
pomeron exchanges, as such they  provide a novel window to pomeron physics in QCD.

With a wide coverage for backward rapidities, single diffractive reactions should be
conveniently selected with the target proton staying intact. Although this physics was 
explored at the Tevatron, experiments at AFTER would allow a wider domain of exploration, 
particularly using nuclear targets.

One can also study the diffractive dissociation of the proton to three jets, 
thus measuring the three-quark valence light-front wavefunction of the projectile 
wavefunction~\cite{Frankfurt:2002jq,Ivanov:2008ax}, in analogy to the E791 measurements of the 
diffractive dissociation of a pion jet~\cite{Aitala:2000hb,Ashery:2006zw}.

This analysis would be 
done effectively by looking at the 3 jets in the target-rapidity region with the absence of a forward activity.
The invariant mass of the  3 jet system can be as high as 30 
GeV. In $pA$, the cross section should scales as $A^2 F^2_A(t)$ where $F_A(t)$ is the 
nuclear form factor~\cite{Frankfurt:2002jq}. Such nuclear dependence could be
studied in Pb$p$ collisions by looking at mini-jets in the target region.

One also can test PQCD color transparency~\cite{Brodsky:1988xz}: the prediction 
that there is no absorption of the initial state proton projectile in hard diffractive 
reactions. Such effect has been observed by E791~\cite{Aitala:2000hc} for $\pi$ projectiles but never for
proton beams.

\subsection{Heavy-hadron (diffractive) production at $x_F \to -1$}

A significant source of charmonia at large --positive-- $x_F$
can be attributed to the projectile-IC coalescence. A similar effect
for light quarks results in the so-called {\it leading-particle} effect. 
Target-IC coalescence should also generate an excess of charmonia in the backward 
region, for say $x_F< -0.1$. The reason is simple: the constituents of a 
given intrinsic heavy-quark Fock state tend to have all the same rapidity, being that
of the projectile or that of the target. From this hypothesis, early $\Lambda_c^+$ data 
at $\sqrt{s}=62$ GeV~\cite{Basile:1981wh} could be accounted for~\cite{Barger:1981sv}.
It is also possible  that the unexpected observations~\cite{Basile:1981nr} of
leading $\Lambda_b$ at high $x_F$ at $\sqrt{s}=62$ GeV could be due the coalescence 
of the $u d b$ constituents of the projectile $|uud b \bar b\rangle$  Fock state.
Other hints for enhanced heavy-hadron production at large $x_F$ are the claimed production
 double-charm $\Xi_{cc}^+$ baryons by SELEX~\cite{Mattson:2002vu,Ocherashvili:2004hi},
difficulty explained otherwise\footnote{And so far unseen in $e^+e^-$ reactions~\cite{Aubert:2006qw}.}, 
and the large-$x_F$ production of $J/\psi$ pairs at NA3~\cite{Badier:1982ae},
consistent with double-IC Fock states~\cite{Vogt:1995tf}. 

All this certainly motivates for modern studies of heavy-hadron production 
at large $|x_F|$ and, even more, for the first to date in the far backward region. 
The first aim would be to confirm such enhancement
of their production cross section, both for charmed and beauty hadrons. For $\Lambda_b$
we could use its decay into $J/\psi \Lambda$ and trigger on $J/\psi$.
If the magnitude of intrinsic heavy-quark Fock states allows it, we could
then look for $\Xi^{++}_{cc}$ to solve the discrepancy
between SELEX and $B$-factories. Next, we could look for the triply heavy 
baryons $\Omega^{++}_{ccc}$ and $\Omega^-_{bbb}$, undiscovered so far. 

\subsection{Very backward physics}

Using deuterium target, it is possible to study the hidden-color excitations of the deuteron~\cite{Matveev:1977xt}. 
There are five distinct color-singlet 
representations of the six quark valence state of the deuteron, only one of which, the 
$n-p$ state, is considered in conventional nuclear physics (for a review see~\cite{Bergstrom:1979fp}). In PQCD all five Fock states 
mix by gluon exchange; at short distances the deuteron distribution amplitude evolves to 
equal admixtures of the five states.  These novel ``hidden-color"  components~\cite{Brodsky:1983vf} 
can be studied at AFTER by probing parton distributions in inclusive reactions requiring high 
$x$ ($ \geq 1 $) and by studying the diffractive dissociation of the deuteron
 in its rapidity domain~\cite{Akhieser:1957zz} in Pb$d$ collisions.

\subsection{Direct hadron production}

AFTER will be able to investigate direct hadron production at high 
transverse momentum~\cite{Arleo:2009ch} where the detected hadron is formed within the hard 
subprocess such as $g q \to \pi q$ and $q q \to p \bar q$, rather than from 
jet fragmentation. Such higher twist color-transparent processes are believed 
to underlie the observed anomalous power law fall-off of inclusive cross section 
for $p p \to HX$  at fixed $x_T$ and $\theta_{CM},$ as well as the anomalously 
large baryon-to-meson ratios seen in central heavy ion collisions.

\section{Further potentialities of a high-energy fixed-target set-up}

\subsection{$D$ and $B$ physics}

Taking advantage of the boost between the center-of-momentum frame and the laboratory, 
$B$ physics studies may also be an important part of a physics case for a fixed
target experiment on the LHC beam. They were in fact the prime motivation
for proposals in the 90's for such a fixed-target set-up on the SSC, the SFT proposal~\cite{Cox:1990jk}, 
and on the LHC, the LHB proposal~\cite{LHB}. 

Much information on $CP$ violation in the $B$ sector has been gathered in the meantime, thanks
to the $B$ factories at SLAC and KEK. It is still being actively studied by the LHCb experiment.
Specific complementary studies of flavour oscillations for instance, using the large $\gamma$-factor --60--
could be envisioned. Following the estimation done for LHB, we expect  to collect possibly up to $10^{10}$ $B$'s
per year. It has also been recently discussed that a fixed-target facility on the Tevatron would allow
one to study $D^0-\bar D^0$ oscillations and to check for possible $CP$ violation in this sector~\cite{Adams:2009mc}.
This would apply here as well. The advantages compared to $B$ factories and LHCb are discussed in~\cite{Adams:2009mc}.

\subsection{Secondary beams}

In principle, TeV secondary beams of $\pi$, $K$ could be created by impinging the extracted 7 TeV beam on a
primary target. The importance of pion-beam facilities has been lately re-emphasized by the $\pi N$ DY COMPASS program~\cite{COMPASS-II} 
to measure the Boer-Mulders functions. In general, meson beams are particularly interesting since they carry a valence antiquark. 
It has also been suggested~\cite{Uggerhoj:2005xz} that $e^+/e^-$  tertiary beams with an energy up to 4 TeV
could be achieved with an efficiency up to $10^{-5}$.

For the sake of completeness, let us mention that, 
despite the requirement of significant extra civil engineering, 
the creation of hundred-GeV neutrino beam might be
possible. This would offer opportunities for $\nu$ DIS at small $x$ (see~\cite{Adams:2009mc}) .

\subsection{Forward studies in relation with cosmic shower}

The main uncertainty in air-shower experiments is currently
associated to the production of muons, and thus the interactions (and
the decay) of $\pi$, $K$ and charmed mesons (see \eg~\cite{Ulrich:2010rg}). Studies of the interactions of
such meson on air-like targets (e.g. carbon, nitrogen) could be
extremely helpful if meson secondary beams could be produced.

In addition, proton-air and even lead-air measurements may also be
very informative since they are not very well constrained by data at higher
energies whereas they still  play some --smaller-- role during the air shower
cascade. Provided that the forward region can be instrumented, such a facility would 
certainly allow one to measure meson and baryon multiplicities and
differential cross sections in proton-air collisions.


\section{Conclusions}
\label{sec:conclusions}

A fixed-target facility based on the multi-TeV proton or heavy ion beams at the 
LHC extracted by a bent crystal, interacting with a fixed proton, deuteron or nuclear 
target, can provide a novel testing ground for QCD at unprecedented  laboratory 
energies and momentum transfers.  Experiments  at negative $x_F$ which detect 
mid and high $p_T$ hadrons and photons emerging from the target-rapidity hemisphere and beyond,
$-4.8 < y_{cms} <  1$,  provide important constraints on the proton and neutron 
valence-quark and gluon dynamics at large $x$ as well as on nuclear effects in $pA$ collisions.  A polarized target  allows 
the study of spin correlations such as the non-factorizing Sivers mechanism in 
Drell-Yan as well as in gluon-sensitive reactions and the surprisingly large single-spin asymmetries observed 
in semi-inclusive reactions, \eg~ $ p p^\uparrow \to \pi X$ at high $x^\uparrow.$  

The LHC heavy ion beam interacting on a variety of nuclear targets allows the 
systematic study of the quark-gluon plasma from the perspective of the target rest 
frame in nucleus-nucleus collisions at center-of-mass energies up to $\sqrtS{NN} = 72$ 
GeV.  In the case of the ion beam colliding on a proton target, one can study the 
diffractive dissociation of the proton into three jets and tests of color transparency.  
The domain $x >1$ in a nuclear target can probe novel aspects of the nuclear wavefunction such 
as hidden color.

A fixed-target facility utilizing the high energy LHC beam leads to the possibility 
of producing very heavy baryons such as $\Omega_{ccc}$, $\Omega_{bbb}$, $\Xi^+_{ccb}$, ...,  
as well as single and double heavy-quark meson production such as $B_c$ 
in diffractive and non-diffractive channels. This production can occur at high Feynman 
momentum fractions, $x_F \sim -1$, in the target-rapidity domain because of the intrinsic 
charm and bottom Fock states of the target.

We also note that such a fixed-target facility at the LHC could produce high energy secondary beams which 
 greatly increases its versatility. It can also provide a valuable testing ground to verify the physics
of high energy air-showers provided that the forward region can be instrumented. Overall, a whole 
spectrum of analyses pertaining to nuclear, hadronic and particle physics
can be covered by such a project. Its cost would be reasonable and its operation 
would not alter at all that of other LHC experiments.


\section*{Acknowledgments}

We are grateful to M.~Anselmino, F.~Arleo, R.~Arnaldi, D. Boer, Z.~Conesa del Valle, D.~d'Enterria, 
J.P.~Didelez, E.G.~Ferreiro,  D.S.~Hwang, L.~Kluberg, C.~Lorc\'e, A.~Rakotozafindrabe, W.~Scandale, I. Schienbein, 
E.~Scomparin,  A.~Sickles, M.~Strikman, U.~I.~Uggerh\o j, T.~Ullrich, R.~Ulrich,
W. Vogelsang and R.~Vogt for useful and stimulating discussions.

This research was supported in part by the Department of Energy, contract DE--AC02--76SF00515 and by
the France-Stanford Center for Interdisciplinary Studies (FSCIS).


\end{document}